\documentclass[floats,floatfix,amssymb,prd,twocolumn,superscriptaddress,nofootinbib, preprintnumbers]{revtex4-1}
	
\usepackage{subcaption}
\usepackage{ragged2e}
\DeclareCaptionJustification{justified}{\justifying}
  \usepackage{orcidlink}
\usepackage{tikz}
\usepackage{amssymb,amsmath,verbatim,mathtools,needspace,enumitem,etoolbox,graphicx,physics,microtype,afterpage,bm}
\usepackage{esint}
\usepackage[all]{hypcap}
\usepackage[T1]{fontenc}
\usepackage[utf8]{inputenc}
\usepackage{tabularx}
\usepackage{float}
\renewcommand{\arraystretch}{1.4}

\newcommand\underrel[3][]{\mathrel{\mathop{#3}\limits_{%
      \ifx c#1\relax\mathclap{#2}\else#2\fi}}}
\usepackage{soul}
\usepackage{lmodern}
\allowdisplaybreaks
\usepackage{tikz}
\usepackage{color}
\usepackage{framed}
\usepackage{hyperref}
\hypersetup{colorlinks, citecolor=bluscuro, linkcolor=black, urlcolor=bluscuro}
\definecolor{rossos}{cmyk}{0,1,1,0.55}
\definecolor{bluscuro}{rgb}{0.15, 0.2, .85}
\definecolor{bluchiaro}{cmyk}{1,.3,0.,0.1}
\definecolor{ForestGreen}{rgb}{0.13, 0.55, 0.13}

\newcommand{\be}{\begin{equation}}
\newcommand{\ee}{\end{equation}}

\newcommand{\llp}{\left [}
\newcommand{\rrp}{\right ]}
\newcommand{\lp}{\left (}
\newcommand{\rp}{\right )}

\def\lsim{\mathrel{\rlap{\lower4pt\hbox{\hskip0.5pt$\sim$}}
    \raise1pt\hbox{$<$}}}         
\def\gsim{\mathrel{\rlap{\lower4pt\hbox{\hskip0.5pt$\sim$}}
    \raise1pt\hbox{$>$}}}         

\makeatletter
\newcommand{\subsetsim}{\mathrel{\mathpalette\subset@sim\relax}}
\newcommand{\subset@sim}[2]{%
  \vtop{\offinterlineskip\m@th
    \ialign{\hfil##\cr
     ~$#1\subset$\cr\noalign{\kern0.5pt}\scalebox{0.9}{$#1\sim$}\cr
    }%
  }%
}
\makeatother

\makeatletter
\def\l@subsubsection#1#2{}
\makeatother

\begin{document}

\title{Flea on the elephant: \\
Tidal Love numbers in subsolar primordial black hole searches}

\author{Valerio De Luca\orcidlink{0000-0002-1444-5372}}
\email{vdeluca@sas.upenn.edu}
\affiliation{Center for Particle Cosmology, Department of Physics and Astronomy,
University of Pennsylvania 209 S. 33rd St., Philadelphia, PA 19104, USA}

\author{Gabriele Franciolini\orcidlink{0000-0002-6892-9145}}
\email{gabriele.franciolini@cern.ch}
\affiliation{CERN, Theoretical Physics Department, Esplanade des Particules 1, Geneva 1211, Switzerland}

\author{Antonio Riotto\orcidlink{0000-0001-6948-0856}}
\email{antonio.riotto@unige.ch}
\affiliation{D\'epartement de Physique Th\'eorique and Gravitational Wave Science Center (GWSC), Universit\'e de Gen\`eve, CH-1211 Geneva, Switzerland}
\affiliation{LAPTh, CNRS USMB F-74940 Annecy, France}


\begin{abstract}
\noindent
Detecting subsolar objects in black hole binary mergers is considered a smoking gun signature of primordial black holes.
Their supposedly vanishing tidal Love number is generically thought to help distinguish them from other subsolar and more deformable compact objects, such as neutron stars.  We show that a large and detectable  Love number of primordial black holes can be generated in the presence of even small disturbances of the system, thus potentially jeopardizing their discovery. However, 
such small perturbations are not tightly bound and are therefore disrupted before the mergers. We show that they leave a characteristic signature in the gravitational waveform that could be observed with current and future gravitational wave detectors. Thus, they may still hint towards the primordial nature of the black holes in the merger.  
Finally, we demonstrate that disregarding possible environmental effects in the matched-filter search for subsolar gravitational wave events can lead to a decreased sensitivity in the detectors.
\end{abstract}

\preprint{ET-0480A-24. CERN-TH-2024-140}
\maketitle

\section{Introduction}
\label{sec:intro}
\noindent
The detection of a subsolar mass object in a black hole (BH) merger is regarded as one of the definitive proofs of the primordial origin of a binary \cite{Green:2020jor}. However, leveraging this opportunity requires the ability to differentiate such an event from other astrophysical systems and possible new physics candidates~\cite{Cardoso:2019rvt,Barsanti:2021ydd,Franciolini:2021xbq,Crescimbeni:2024cwh}.

Subsolar compact objects may have astrophysical origin, e.g. white dwarfs  \cite{Kilic:2006as} and neutron stars  \cite{Lattimer:2004pg,Metzger:2024ujc}, but they could also exist within the realm of beyond-Standard-Model physics~\cite{Giudice:2016zpa,Cardoso:2019rvt} and might have a cosmological origin. 
Examples include Q-balls~\cite{Coleman:1985ki}, boson stars~\cite{Liebling:2012fv}, and fermion-soliton stars~\cite{Lee:1986tr,DelGrosso:2023trq,DelGrosso:2023dmv, Berti:2024moe} (see~\cite{Cardoso:2019rvt} for an overview). Many of these models can accommodate subsolar compact objects, depending on the coupling of the underlying fundamental theory~\cite{DelGrosso:2024wmy}.

A key distinction between material compact objects and BHs is that the tidal deformability -- measured by tidal Love numbers (TLNs) --  of the latter is (supposed to be) zero in four-dimensional space-times~\cite{Deruelle:1984hq,Binnington:2009bb, Damour:2009vw, Damour:2009va, Pani:2015hfa, Pani:2015nua, Gurlebeck:2015xpa, Porto:2016zng, LeTiec:2020spy, Chia:2020yla, LeTiec:2020bos, Hui:2020xxx,Charalambous:2021mea,  Charalambous:2021kcz, Creci:2021rkz,  Bonelli:2021uvf, Ivanov:2022hlo, Charalambous:2022rre, Katagiri:2022vyz, Ivanov:2022qqt, Berens:2022ebl, Bhatt:2023zsy, Sharma:2024hlz, Rai:2024lho}. This unique property of BHs in general relativity can be explained by special symmetries of the equations of motion for static and linear perturbations~\cite{Hui:2020xxx,Charalambous:2021mea,Charalambous:2021kcz,Hui:2021vcv,Hui:2022vbh,Charalambous:2022rre,Ivanov:2022qqt,Katagiri:2022vyz, Bonelli:2021uvf,Kehagias:2022ndy,BenAchour:2022uqo,Berens:2022ebl, Rai:2024lho,Kehagias:2024yzn, Sharma:2024hlz} (see also Refs.~\cite{DeLuca:2023mio, Riva:2023rcm, Ivanov:2024sds, Nair:2022xfm, Saketh:2023bul, Perry:2023wmm, Chakraborty:2023zed, DeLuca:2024ufn} for recent studies discussing the role of nonlinearities and time-dependent tidal fields), but it does not apply to any other material object~\cite{Porto:2016zng, Cardoso:2017cfl}.

This property is however invalidated in scenarios involving the presence of a cosmological constant~\cite{Nair:2024mya}, in theories of modified gravity~\cite{Cardoso:2017cfl, Cardoso:2018ptl, DeLuca:2022tkm, Barura:2024uog} or higher dimensions~\cite{Kol:2011vg, Cardoso:2019vof, Hui:2020xxx, Rodriguez:2023xjd, Charalambous:2023jgq, Charalambous:2024tdj, Charalambous:2024gpf, Ma:2024few}, and especially in the presence of an external environment. The latter significantly alters the nature of the tidal interactions, potentially leading to nonzero TLNs for nonvacuum binary BHs. Examples of this effect include BHs surrounded by clouds of ultralight bosonic fields generated through accretion or superradiance~\cite{Baumann:2018vus, DeLuca:2021ite, DeLuca:2022xlz, Brito:2023pyl}, or by fluids of matter and accretion disks~\cite{Cardoso:2019upw, Cardoso:2021wlq,Cannizzaro:2024fpz}.

Since TLNs affect the gravitational wave (GW) signal of binary systems at high post-Newtonian (PN) order~\cite{Flanagan:2007ix,Hinderer:2009ca} and could be measured with good accuracy in the presence of large tidal interactions~\cite{Chia:2023tle, Chia:2024bwc}, they provide a natural key feature to test the nature of the compact object.

Suppose now that in the future a subsolar merger event is detected with a large TLN. One would be immediately driven to think that such an event does not involve a primordial black hole (PBH).   We show in this paper that this is not the case. Indeed, 
 even a small environmental disturbance (the ``flea'') to a BH (the ``elephant'') generates a large and measurable effective TLN of the system.  
Such a  ``flea on the elephant'' effect,
here named in analogy with the quantum mechanical effect~\cite{Graffi_1984},
is not surprising, it appears as well in the 
 spectrum of the quasinormal modes of the GWs emitted during the ringdown
phase following the merger of two BHs when tiny perturbations are present \cite{Jaramillo:2020tuu,Destounis:2021lum,Cheung:2021bol, Gasperin:2021kfv,Boyanov:2022ark,Jaramillo:2022kuv,Kyutoku:2022gbr,Sarkar:2023rhp,Destounis:2023nmb,Arean:2023ejh,Cownden:2023dam,Destounis:2023ruj,Courty:2023rxk,Boyanov:2023qqf,Cao:2024oud,Cardoso:2024mrw,Ianniccari:2024ysv}. 

This raises a crucial question: is tidal deformability a robust observable to disentangle PBHs from other compact objects? We show that the answer is still yes. Indeed, the tiny disturbances that generate a large TLN for the PBHs are typically disrupted by tidal interactions. This phenomenon leaves a characteristic imprint in the gravitational waveform which will still allow us to identify PBH players in the BH mergers. 
Furthermore, we will show that neglecting these effects in the matched-filter search for subsolar GW events in the data can cause a reduction in sensitivity.

The paper is organized as follows. In Sec. II we
outline the computation of the effective TLN for the ``flea on the elephant'' system. In Sec. III we discuss the prospects of detectability and measurability for subsolar PBH binaries at present and future GW experiments. In Sec. IV we present the main conclusions of
our work. In the following we use geometrical units $c = G = 1$.

\section{An Effective tidal Love number}
\label{sec: TLN}
\noindent 
The TLN provides information about  the  structure of the deformed compact object and -- 
interestingly enough -- it vanishes for non-rotating and spinning BHs in the static limit~\cite{Binnington:2009bb,Damour:2009vw,Damour:2009va,Pani:2015hfa,Pani:2015nua,Gurlebeck:2015xpa,Porto:2016zng,LeTiec:2020spy, Chia:2020yla,LeTiec:2020bos}, due to underlying hidden symmetries of general relativity (at least at the linear level).

A straightforward way to find the linear TLN in general relativity is to solve for the dynamics of the helicity two tensor degrees of freedom $h_\ell$ (expanded in a spherical harmonics decomposition with multipole $\ell$)  in the static regime. At infinity, for asymptotically flat space-times, the general solution has a   growing mode $\sim r^{\ell+1}$,  which represents the tidal force by which the reaction of the spacetime metric describing the body (in our case the BH) is tested. The reaction is measured by the decaying mode $\sim r^{-\ell}$. 
The TLN $k_\ell$ is fixed by the ratio of the  coefficients of the decaying and growing modes, in analogy to the asymptotic expansion of the gravitational potential in Newtonian gravity, as~\cite{Binnington:2009bb, Damour:2009vw, Damour:2009va}
\begin{equation}
\label{TlNexp}
h_\ell (r) \propto r^{\ell+1} \left[ 1 + \cdots  + k_\ell \left( \frac{r_s}{r} \right)^{2\ell +1} + \cdots  \right]\,,
\end{equation}
once the solution satisfies the condition of regularity at the BH horizon $r_s$. For non-rotating and spinning BHs in the vacuum the decaying mode $r^{-\ell}$ is mapped into the divergent solution at the horizon and it must be excluded from the set of physical solutions, hence the vanishing of the TLN.

As a working example to find the effect of 
 a small disturbance to the TLN, 
we consider the equation of motion of stationary tensor modes in a Schwarzschild spacetime when a small potential bump is added away from the horizon~\cite{Barausse:2014tra} 
\begin{equation}
\label{eqs}
\left[\partial_{r_*}^2-V^\epsilon_\ell(r)\right] h_{\ell}(r)=0\,,  
\end{equation}
where 
 \be
r_*=r+r_s\ln\left|\frac{r-r_s}{r_s}\right|
 \ee
is the standard tortoise coordinate and $r_s$ is the Schwarzschild radius (for a spherically symmetric background, the azimuthal number is degenerate and can be set to zero without loss of generality). 

\begin{figure}[t!]
    \centering
    \includegraphics[width=0.5\textwidth]{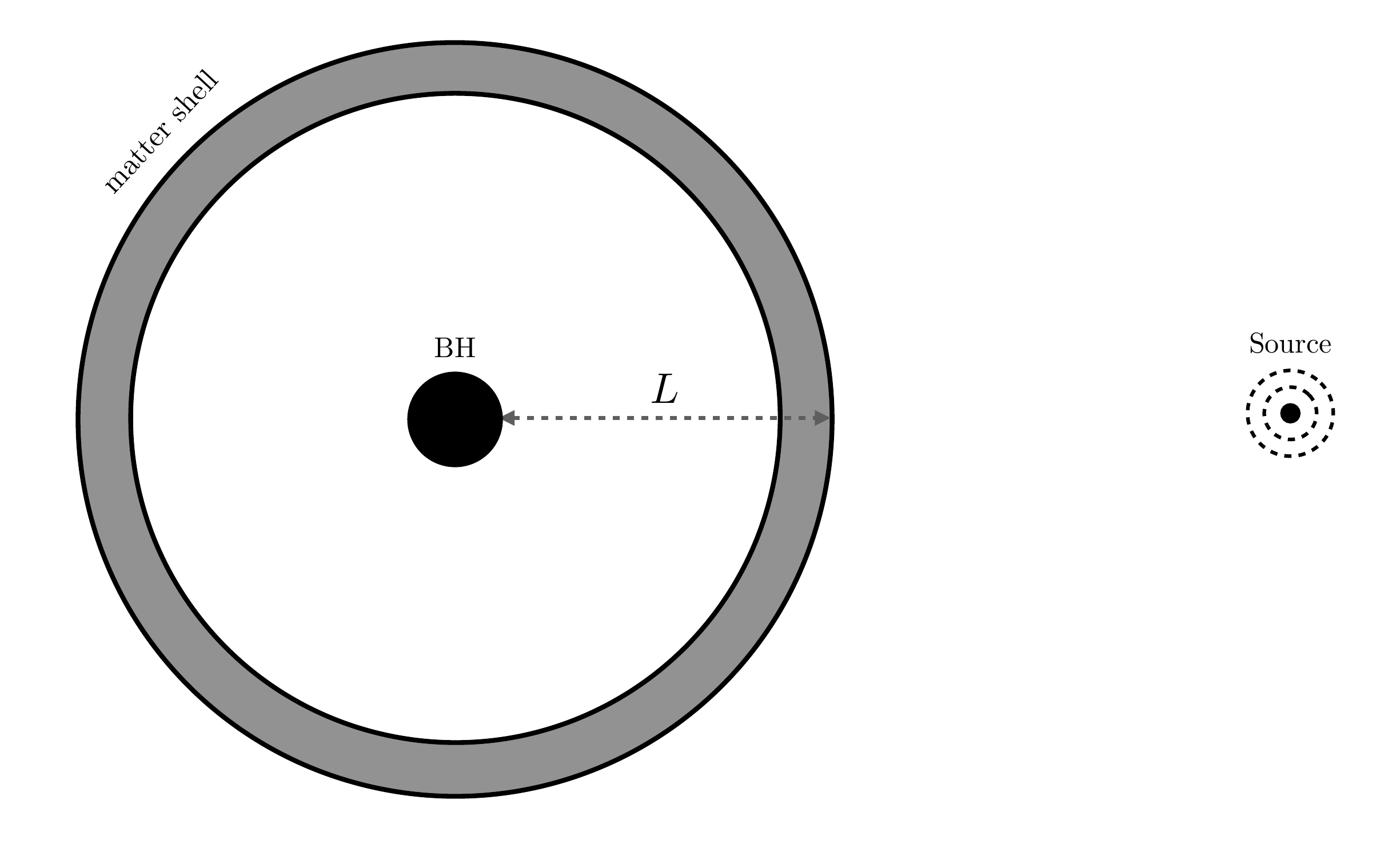}
    \caption{ 
Schematic representation of the configuration of a thin matter shell around the BH and the source generating an effective tidal Love number.}\label{fig:1}
\end{figure}

The potential reads (say, for the odd parity helicity)~\cite{Cheung:2021bol}
\begin{align}
\label{eq:potentials}
V^\epsilon_\ell(r)&=V_0(r)+V_1(r)\,,
\nonumber \\
V_0(r)&=\left(1-\frac{r_s}{r}\right)\left[\frac{\ell(\ell+1)}{r^2}-\frac{3r_s}{r^3}\right]\,,\nonumber\\
V_1&=
\frac{\epsilon}{r_s}\,\delta(r_*-L)\,. 
\end{align}
The Dirac delta potential $V_1$, describing the external bump, is located at a distance $L$ much larger than $r_s$, and is the same considered in the  aforementioned literature regarding the quasinormal modes~\cite{Jaramillo:2020tuu,Destounis:2021lum,Cheung:2021bol, Gasperin:2021kfv,Boyanov:2022ark,Jaramillo:2022kuv,Kyutoku:2022gbr,Sarkar:2023rhp,Destounis:2023nmb,Arean:2023ejh,Cownden:2023dam,Destounis:2023ruj,Courty:2023rxk,Boyanov:2023qqf,Cao:2024oud,Cardoso:2024mrw,Ianniccari:2024ysv}, sharing strong similarities with the two rectangular barriers model of Refs.~\cite{Chandrasekhar:1975zza, Barausse:2014tra, Cheung:2021bol}.
It is supposed to mimic the presence of a matter shell around the BH, with density profile characterised by a Poschl-Teller-like peak~\cite{Cheung:2021bol}.  The density amplitude is  ${\cal O}(\epsilon)$,  with width ${\cal O}(r_s)\ll L$ and corresponding mass  $\sim \epsilon \lp L/r_s\rp m$, where $m$ is the BH mass. 
 This simple toy model allows to have an admittedly partial, but good, description of environmental effects present around the BH system~\cite{Baumann:2018vus, DeLuca:2021ite, DeLuca:2022xlz, Brito:2023pyl,Cardoso:2019upw, Cardoso:2021wlq,Cannizzaro:2024fpz},  and has provided valuable insights on their role for BH spectroscopy, at least from an analytical perspective. It mimics for example a layer of a more complicated matter distribution which might spread  over larger distances. The total effect of such matter distribution could thus be described as an integrated effect over many shells of this form. Our results are therefore intended to be conservative as far as the TLN is concerned, since tidal interactions could be effective for a longer portion of the inspiral (see discussion of tidal disruption below). The source creating the tidal force is assumed to be at distances $\gg L$, see Fig. \ref{fig:1} for a pictorial illustration.

To determine the TLN, we will follow a perturbative approach in the coupling $\epsilon$, demanding regularity of the solution at the BH horizon, at each order in perturbation theory. At zeroth and first order in $\epsilon$, the equation of motion for an axial static perturbation reads
\begin{align}
& (\partial_{r_*}^2 - V_0 (r)) h_\ell^{(0)} = 0\,, \nonumber \\
& (\partial_{r_*}^2 - V_0 (r)) h_\ell^{(1)} = \epsilon V_1 (r) h_\ell^{(0)} \,.
\end{align}
The regular solution at the leading  order, for the physically relevant $\ell = 2$ mode, reads
\begin{align}
h_2^{(0)} = c_1 \left( \frac{r}{r_s} \right)^3\,,     \end{align}
where $c_1$ denotes the strength of the tidal term. By requiring regularity at the horizon, 
the final solution up to the first order in $\epsilon$ reads
\begin{widetext}
\begin{align}
h_2 = h_2^{(0)} + h_2^{(1)} & =
c_1 \lp \frac{r}{r_s} \rp^3 
+ c_1
\frac{\epsilon L^3  }{12 r_s^5 r (L-r_s)}  \Theta (r-L) \left\{12 L^4 \lp \frac{r}{r_s} \rp^4 \ln \left|\frac{L(r-r_s)}{r(L-r_s)}  \right| +12 L^4 \lp \frac{r}{r_s} \rp^3 \right. \nonumber \\
& \left. +6 L^4 \lp \frac{r}{r_s} \rp^2+4 L^4 \lp \frac{r}{r_s} \rp+3 L^4-r_s \lp \frac{r}{r_s} \rp^4 \left(12 L^3+6 L^2 r_s+4 L r_s^2+3 r_s^3\right)\right\}\,,
\end{align}
\end{widetext}
where $\Theta$ indicates the Heaviside step function. Notice that the first-order term is proportional to the tidal field strength $c_1$, as expected in a perturbative approach.
By expanding the solution at spatial infinity one gets 
\begin{align}
h_2(r\gg L) \approx c_1 \left(\frac{r}{r_s}\right)^3\left[1-\frac{\epsilon}{5}\left(\frac{L}{r_s}\right)^6\left(\frac{r_s}{r}\right)^5\right]\,,
\end{align}
where the growing mode describes the tidal potential and the decaying mode characterizes the system response.
The corresponding TLN can be derived from the relative coefficient between the two terms, as shown in Eq.~\eqref{TlNexp}, and reads \cite{poisson_will_2014}
\begin{equation}\label{eq:TLN}
k_2 = - \frac{\epsilon}{5} \left( \frac{L}{r_s} \right)^6.
\end{equation}
Thus, the  TLNs of PBHs surrounded by weak rings could be sufficiently large
and potentially impact the evolution of coalescing binaries. To get the feeling of the numbers, for the characteristic values $\epsilon\sim 10^{-4}$ and $L\sim 30\, r_s$, one obtains sizeable TLNs as $k_2\sim - 1.5\cdot 10^4$. This shows that the presence of an external bump, induced by environmental effect, would switch on potentially large tidal effects in the inspiral part of the signal.  
It is worth commenting that a similar computation has been performed in
Ref.~\cite{Katagiri:2023yzm} for a Gaussian small bump. The size and dependence of the corresponding TLNs on the bump parameters are compatible with ours, when overlap is possible.

However, in the final inspiral phases, the environment around each PBH may be disrupted because of tidal interactions. This disruption typically occurs when the binary's semi-major axis is comparable to the Roche radius $R_\text{\tiny Roche}$ of the system~\cite{Shapiro:1983du}, which is defined as the distance at which the matter shell, held together solely by its gravitational force $F_g$ to the BH, will break apart due to the tidal force $F_t$ of a companion. This happens when the corresponding forces, acting on a volume element of mass $u$, are equal to each other, which is
\begin{equation}
F_g \simeq \frac{m_1 u }{(r_{s,1} + L)^2} = F_t \simeq \frac{m_2 (r_{s,1} + L) u }{R_\text{\tiny Roche}^3}\,.
\end{equation}
When the two masses are equal, one naturally obtains $R_\text{\tiny Roche}\simeq L$.
In this estimate, we have considered an idealized scenario where the secondary BH's environment is negligible compared to the secondary BH itself, because of its low density, destroying the ring surrounding its companion much before the frequency corresponding to the innermost stable circular orbit (ISCO), $f_\text{\tiny ISCO} \simeq 4.4 \, {\rm kHz}\, M_\odot/(m_1+m_2)$, is reached. 

The GW frequency associated with tidal disruption for a circular binary then takes the form
\begin{align}
\label{fcut}
f_\text{\tiny r} &\simeq \frac{1}{2 \sqrt{2} \pi  m_1} \sqrt{\frac{(m_1+m_2)}{(1+\tilde{L})^3 m_2}}\,
\nonumber \\
& \simeq 393 \, {\rm Hz} \lp\frac{m_1}{0.5 M_\odot} \rp^{-1} \lp\frac{\tilde L }{30} \rp^{-3/2}\,,
\end{align}
where we have introduced the rescaled bump distance $\tilde{L}\equiv L/r_{s,1}$  and in the last step we have focused on equal mass binaries $m_2 = m_1$ to provide a simple estimate. 
The system is therefore described by a frequency-dependent TLN~\cite{DeLuca:2021ite, DeLuca:2022xlz}, which is sizeable and nonvanishing in the early inspiral and dies off smoothly after the Roche frequency, leaving a system of two naked BHs.

PBHs with larger  $\tilde{L}$, or smaller amplitude $\epsilon$,  are disrupted earlier. While a small $\epsilon$ reduces the relevance of tidal effects, a small $\tilde{L}$ makes TLNs smaller but their impact lasts longer. The interplay between these factors will be crucial to address their detectability in GW experiments, as we will show in the next section.

\section{Observational prospects}
\label{sec: Fisher}
\noindent
We estimate the observational prospects of measuring tidal effects for subsolar binaries at current and future GW detectors using the Fisher information matrix (FIM) approach, 
and discuss the relevance of taking into account potential environmental effects when performing a matched-filter search in the data to avoid a reduction in sensitivity of the detectors.

\subsection{Simulations and methods}
\noindent
The FIM~\cite{Poisson:1995ef, Vallisneri:2007ev, Cardoso:2017cfl} relies on the fact that for strong signals with a high signal-to-noise ratio (SNR), the posterior distribution of some model parameters $\vec{\theta}$ can be approximated by a multivariate Gaussian distribution centered at the true values $\vec{\hat{\theta}}$. The covariance matrix of this distribution is given by ${\Sigma}_{ij} = ({\Gamma_{ij}})^{-1}$, where
\be\label{eq:fisher_def}
\Gamma_{ij}= \left\langle \frac{\partial h}{\partial \theta_i}\bigg\vert\frac{\partial h}{\partial \theta_j}\right\rangle_{\vec{\theta}=\vec{\hat{\theta}}}
\ee
is the FIM. The statistical uncertainty in the $i$-th parameter is given by $\sigma_i = \Sigma^{1/2}_{ii}$.
In Eq.~(\ref{eq:fisher_def}), we defined the scalar product of two waveform templates $h_{1,2}$ over the detector noise spectral density $S_n(f)$ as
\be
\langle h_1\vert 
h_2\rangle=4\Re\int_{f_\text{\tiny min}}^{f_\text{\tiny max}} \frac{\tilde{h}_1(f)\tilde{h}^*_2(f)}{S_n(f)}{\rm d}f \,,
\label{scalprod}
\ee
where $*$ denotes complex conjugation, and ${\rm SNR} = \langle h \vert h \rangle^{1/2}$. 
We adopt the minimum and maximum frequencies of integration $f_\text{\tiny min}=10 (1) \, \text{Hz}$ and $f_\text{\tiny max}=f_\text{\tiny ISCO}$, respectively, when considering the fourth run of the LIGO-Virgo-KAGRA collaboration (LVK O4) and the next generation GW detector Einstein Telescope (ET).

To model the GW signal emitted by a binary system characterized by fading environmental effects, such as those associated with the system we discussed in the previous section, we use the standard TaylorF2 waveform~\cite{Damour:2000zb}, with the inclusion of frequency-dependent tidal effects in the GW phase $\psi_\text{\tiny Tidal} (f)$~\cite{DeLuca:2022xlz}. 
In Fourier space, such a signal is given by~\cite{Sathyaprakash:1991mt, Damour:2000gg}:
\be
\tilde h (f) = C_\Omega{\cal A} \, e^{{\rm i} \psi_\text{\tiny PP} (f) + {\rm i} \psi_\text{\tiny Tidal} (f)}\,,
\ee
where $\psi_\text{\tiny PP}$ includes the point-particle terms up to the 3.5PN order~\cite{Damour:2000gg, Arun:2004hn, Buonanno:2009zt, Abdelsalhin:2018reg} and depends on the binary's chirp mass $\mathcal{M} = (m_1 m_2)^{3/5}/(m_1+m_2)^{1/5}$, the symmetric mass ratio $\eta = m_1 m_2 / (m_1+m_2)^2$, where $m_{1,2}$ are the component masses, and the (anti)symmetric combinations of the individual spin components: $\chi_{s} = (\chi_{1} + \chi_{2})/2$ and $\chi_{a} = (\chi_{1} - \chi_{2})/2$. For this analysis, we will assume non-spinning binaries ($\chi_1 = \chi_2 = 0$), as we expect for subsolar PBHs born in a radiation-dominated universe~\cite{DeLuca:2019buf, Mirbabayi:2019uph} which did not experience an efficient phase of baryonic accretion~\cite{DeLuca:2020bjf, DeLuca:2020qqa, DeLuca:2023bcr}, although the spin magnitudes are included in the waveform hyperparameters.
Finally, the leading term in the waveform amplitude is
\be
\mathcal{A}= \sqrt{\frac{5}{24}} 
\frac{\mathcal{M}^{5/6}f^{-7/6}}{\pi^{2/3}d_L} 
\,,
\ee
where $d_L$ is the luminosity distance. The geometric coefficient $C_{\Omega}=[F_+^2 (1+\cos^2 \iota)^2 + 4 F_\times^2 \cos \iota]^{1/2}$ depends on the inclination angle $\iota$ between the binary's line of sight and its orbital angular momentum, as well as on the detector's antenna pattern functions $F_{+,\times} (\theta, \varphi, \psi)$, which are functions of the source's position in the sky $(\theta, \varphi)$ and the polarization angle $\psi$.

The leading tidal correction to the gravitational waveform appears starting from the 5PN order and, taking into account the potential tidal disruption during the binary's evolution, reads
\be
\label{TLN5PN}
\psi_\text{\tiny Tidal} (f) = -\frac{39}{2} (\pi M f)^{5/3}
\tilde{\Lambda} (f) \,,
\ee
where $M = m_1 + m_2$ is the total mass of the binary and $\tilde{\Lambda}(f)$ represents the frequency-dependent effective tidal deformability parameter, which depends on the masses and TLNs of each binary component~\cite{Flanagan:2007ix, Vines:2011ud}. The frequency-dependent effective TLN can be modeled as \cite{DeLuca:2022xlz}
\be
\tilde{\Lambda}(f) \equiv \tilde{\Lambda}_0
\llp \frac{1+e^{-f_\text{\tiny r}/f_\text{\tiny slope}}}{1+e^{(f-f_\text{\tiny r})/f_\text{\tiny slope}}} \rrp,
\ee
and allows tracking a gradual decrease of tidal effects starting from the characteristic Roche frequency $f_\text{\tiny r}$, as indicated in Eq.~\eqref{fcut}, with a characteristic slope $f_\text{\tiny slope}$. Since the slope $f_\text{\tiny slope}$ has minimal impact on the results of the analysis, we will treat it as a fixed parameter rather than a model hyperparameter, setting it to $f_\text{\tiny slope} = f_\text{\tiny r}/5$~\cite{DeLuca:2022xlz}.
The value of $\tilde \Lambda_0$ is derived from the 5PN term as a function of the TLN as~\cite{Flanagan:2007ix, Vines:2011ud}
\begin{align}
\label{LT}
\tilde{\Lambda}_0
&= 
\frac{8}{13}\left[\left(1+7\eta-31\eta^2\right)
\frac{2}{3}
\left(k_{2,1}+k_{2,2}\right) \right .
\nonumber \\
&\left . +\sqrt{1-4\eta}\left(1+9\eta-11\eta^2\right)
\frac{2}{3}
\left(k_{2,1}-k_{2,2}\right) 
\right]\,.
\end{align}
Here $k_{2,i}$ ($i=1,2$) represent the individual TLNs of the two BHs in the binary.
Assuming nearly equal mass objects with similar matter shells around them, one simply finds $\tilde \Lambda_0 \sim 2/3 k_2$. 

Overall, the waveform model we adopt depends on eight 
parameters $\vec{\theta}=\{{\cal M},\eta,\chi_s,\chi_a,t_c,\phi_c,
\tilde \Lambda, f_\text{\tiny r}\}$, 
where $(t_c,\phi_c)$ are 
the coalescence time and phase, set to $t_c=\phi_c=0$, and we have focused on optimally oriented binaries to eliminate the four angular parameters. The results of the Fisher forecast for these hyperparameters allow eventually to get information also on the forecasted uncertainties on both the individual BH masses $(m_1,m_2)$ and on the bump model parameters $(\epsilon, \tilde L)$, as we will show in Sec.~\ref{secmeasur}.

\begin{figure*}[t!]
\centering
\includegraphics[width=0.49\textwidth]{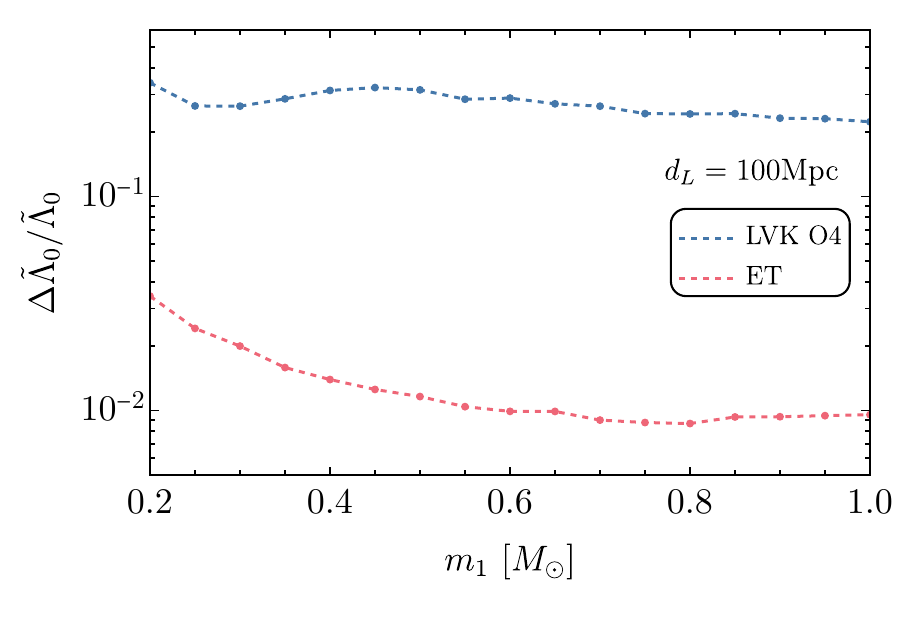}
\includegraphics[width=0.49\textwidth]{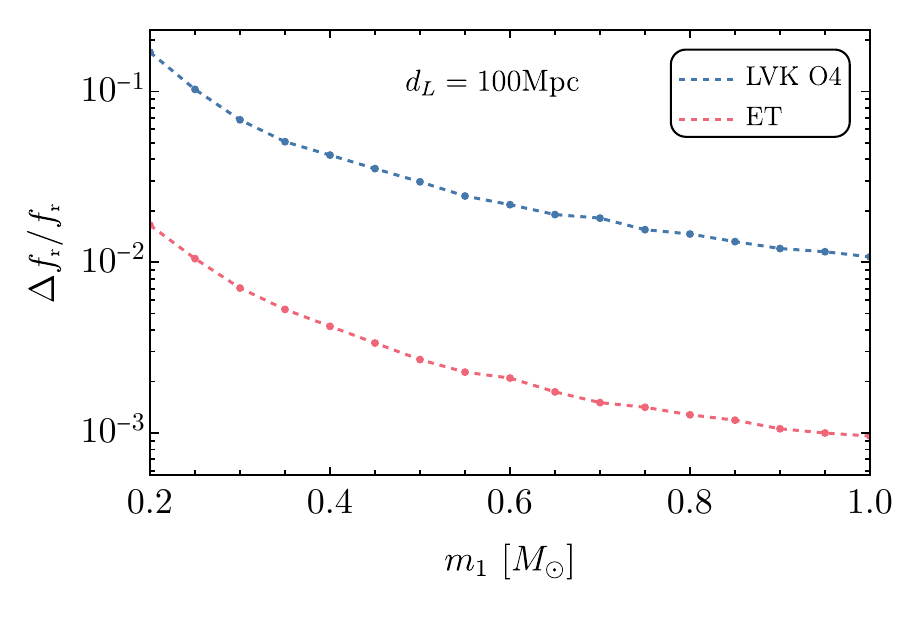}
\caption{ 
Relative uncertainty on the parameters controlling the tidal deformation effects induced by the matter shell around each BH.
We simulate a binary BH located at a distance $d_L = 100 \, {\rm Mpc}$ with optimal orientation and subsolar masses (related by a representative mass ratio $q = 2/3$).
We further assume the disturbance is the same in both systems and we fix $\epsilon =10^{-3}$, $\tilde L = 30$.
The blue lines report the results obtained with current LVK O4 sensitivity, while in red are the future ET one. 
{\it Left panel:}
relative $1\sigma$ uncertainty on the tidal deformability as a function of primary mass.
{\it Right panel:}
relative $1\sigma$ uncertainty on the cut-off frequency $f_\text{\tiny r}$ where deformability effects are switched off, as a function of $m_1$.
}\label{fig:errorLambda0}
\end{figure*}

Furthermore, it is important to stress that neglecting environmental effects in the analyses of subsolar PBH mergers may potentially cause a wrong interpretation of the considered binary event, due to the mismatch between the vacuum and environmental waveforms discussed above. This difference demands the use of the correct waveform when searching for compact binary coalescences using matched filtering against pre-constructed template banks, to avoid a reduction of detector sensitivity (see e.g. 
Ref.~\cite{Wang:2023qgw} for a similar analysis in the context of light and highly spinning neutron star binaries, and Ref.~\cite{Bandopadhyay:2022tbi} for subsolar neutron star mergers).
Furthermore, waveform mismatch would also introduce biases in the binary parameters when performing parameter estimation. 

To assess how much two waveforms differ, we will introduce the faithfulness, defined as the noise-weighted overlap between two normalised templates as~\cite{Owen:1998dk}
\begin{equation}
\mathcal{F} = \max_{\{t_c,\phi_c\}} \frac{\langle h_1\vert 
h_2\rangle}{\sqrt{\langle h_1\vert 
h_1\rangle \langle h_2\vert 
h_2\rangle}}\,,
\end{equation}
where one maximises over the extrinsic parameters of the system, which are the coalescence time $t_c$ and phase $\phi_c$~\cite{Owen:1998dk}.

To get an intuition about the numbers involved, one can assume that values of $\mathcal{F}$ smaller than $\sim 1 - d/(2 \, {\rm SNR}^2)$, where $d$ is the dimension of the waveform model, highlight that two templates differ significantly from each other~\cite{Lindblom:2008cm,Hughes:2018qxz}. 
Assuming a detection with SNR $\sim 12$, and choosing a dimensionality of $d= 8$ in the parameter space (being conservative since the naked model has two parameters less), two signals are distinguishable if $\mathcal{F} \lesssim \mathcal{F}_\text{\tiny th} \simeq 0.972$.
In Sec.~\ref{sec:3C} we will apply the faithfulness criterion to discuss the mismatch between GW waveforms of naked and environmental PBHs.

\subsection{Measurability}
\label{secmeasur}
\noindent
In this section, we forecast the ability of current and future detectors to determine the presence of a small disturbance to the GW signal from the BH environment.

{
\renewcommand{\arraystretch}{1.4}
\setlength{\tabcolsep}{4pt}
\begin{table}[t!]
\begin{tabularx}{\columnwidth}{|X|c|c|}
\hline
\hline
Network  & LVK O4 & ET \\
\hline
\hline
$\Delta m_1/m_1$& 0.17& 0.0069
\\
\hline
$\Delta m_2/m_2$& 0.44&0.017
\\
\hline
$\Delta \tilde \Lambda_0/\tilde \Lambda_0$ & 0.23 & 0.010
\\
\hline
$\Delta f_\text{\tiny r}/f_\text{\tiny r}$ & 0.031&0.0020
\\
\hline
\hline
$\Delta \epsilon/\epsilon$ & 0.26 & 0.014
\\
\hline
$\Delta \tilde L/ \tilde L$ & 0.016 & 0.0014
\\
\hline
\hline
\end{tabularx}
\caption{ 
Standard deviation uncertainty obtained on the masses and deformability parameters assuming the ongoing LVK O4 sensitivity, as well as future ET sensitivity. We simulate a binary system with $m_1 = 0.6 M_\odot$, $m_2 = 0.4 M_\odot$, $\epsilon=10^{-3}$, 
$\tilde L = 30$. We assume a binary at $d_L = 100\ {\rm Mpc}$, which corresponds to SNR = 12.9 (LVK O4) and SNR = 215 (ET).
}
\label{tab:res}
\end{table}
}

\begin{figure*}[t!]
\centering
\includegraphics[width=0.495\textwidth]{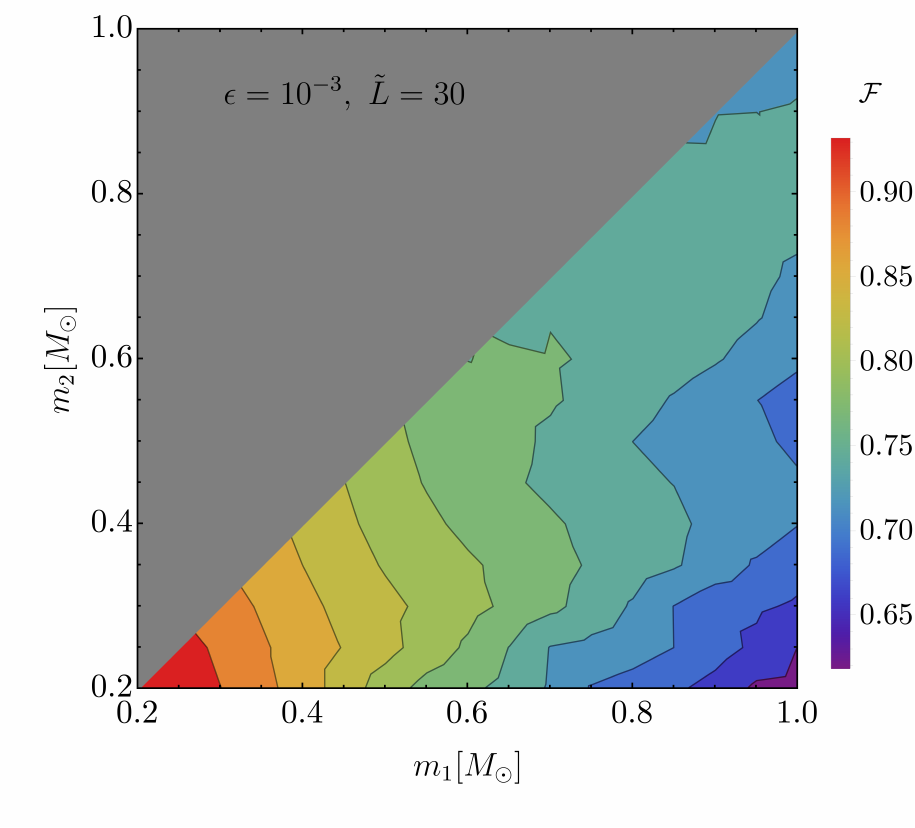}
\includegraphics[width=0.485\textwidth]{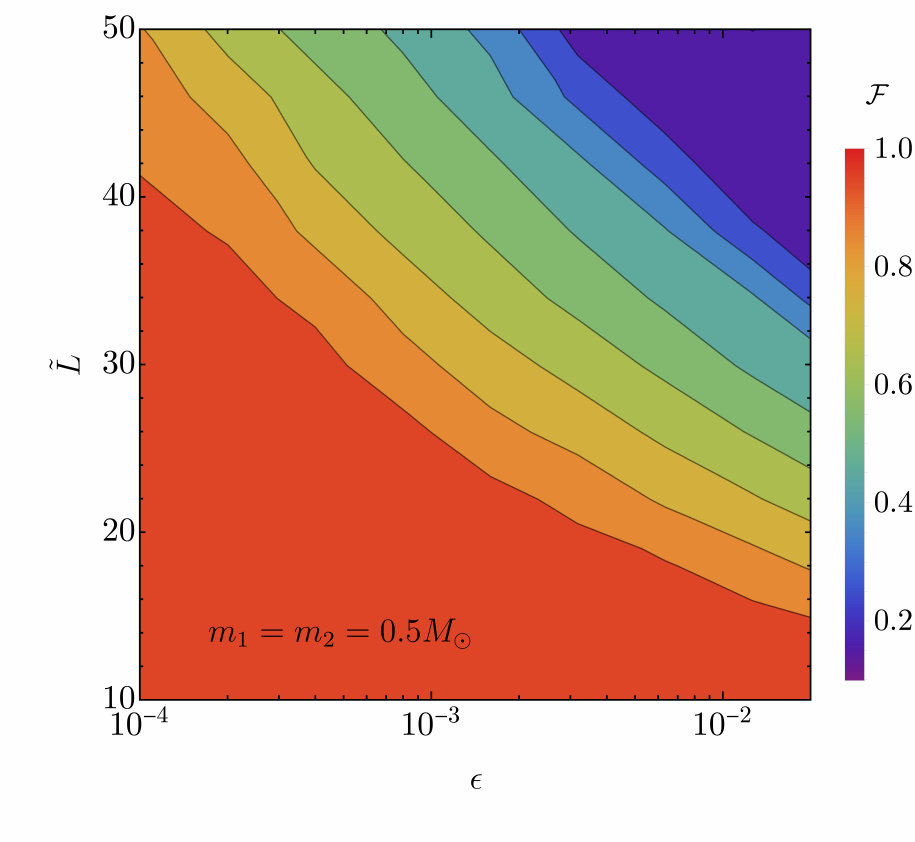}
\caption{ 
Faithfulness $\mathcal{F}$ between GW signals associated to a subsolar PBH binary evolving either in the vacuum or in an environment, parametrised by the thin matter shell. We assume that the binary is observed at ET. 
{\it Left panel:} behaviour in terms of the binary masses, fixing the matter shell parameters to $\epsilon = 10^{-3}$ and $\tilde{L}=30$.
{\it Right panel:} behaviour in terms of the matter shell parameters, fixing the PBH masses to $m_1 =  m_2 = 0.5 M_\odot$.
}\label{fig:Faith}
\end{figure*}

In Fig.~\ref{fig:errorLambda0}, we show how the relative uncertainties on $\tilde \Lambda_0$ and $f_\text{\tiny r}$ scale as a function of the simulated binary primary BH mass $m_1$, for fixed binary distance with respect to the GW ground-based detector. 
For presentation purposes, we fix $\epsilon =10^{-3}$ and $\tilde L = 30$, which corresponds to $\tilde \Lambda_0= 1.1 \times 10^5 $ and $f_\text{\tiny r} = 0.081 f_\text{\tiny ISCO}$ for a representative binary with $m_1 =0.6 M_\odot$ and $m_2 =0.4 M_\odot$.
In blue, we indicate the results obtained with the current LVK O4 sensitivity. Given the limited SNR for the assumed system at $d_L = 100 \, {\rm Mpc}$, which is found to be around the threshold for detection $\text{SNR} \sim 20 \times (m_1/M_\odot)^{5/6}$, the relative uncertainty on $\tilde \Lambda_0$ would fall around $\sim 30\%$, which is close to the threshold for detection with $3\sigma$ credibility. We also see that the uncertainty stops improving when going towards larger masses within the simulated range, as the positive trend in SNR is opposed by the smaller $f_\text{\tiny r}\sim 1/m_1$, therefore reducing the observable information. 
A similar trend is observed for ET, although with a much larger precision due to the larger $\text{SNR} \sim 330\times ( m_1/M_\odot)^{5/6}$, leading to nearly percent uncertainties.

In the right panel of Fig.~\ref{fig:errorLambda0}, we show the relative uncertainties on the cut-off frequency $f_\text{\tiny r} $ where tidal effects switch off. 
Uncertainties are typically smaller than $\sim 10\%$, scaling strongly with the inverse of the primary mass as $\Delta f_\text{\tiny r}/f_\text{\tiny r} \sim {m_1}^{-(5/6+1)}$.
This scaling can be explained by recalling the increasing SNR going towards larger masses, and $f_\text{\tiny r}\sim 1/{m_1}$ for fixed mass ratio. 
The uncertainty reaches percent precision already with LVK O4 sensitivity and $m_1 \gtrsim 0.3 M_\odot$. 
With the ET sensitivity, the uncertainties are reduced by one order of magnitude. This confirms that, although disturbances of the BH environment could induce a large tidal deformability, one could constrain their presence by searching for a cut-off frequency at frequencies smaller than the ISCO.

One can also infer the parameters of the matter shell generating the disturbance from measurements of $(\tilde \Lambda_0, f_\text{\tiny r})$ using Eqs.~\eqref{eq:TLN} and \eqref{fcut}. 
We show this explicitly for a simulated signal in Tab.~\ref{tab:res}. 
It is important to notice here that the TLN scales linearly with $\epsilon$ and with the sixth power of $\tilde L$.
Therefore, as we can observe in Tab.~\ref{tab:res}, relative uncertainties on $\epsilon$ are of similar magnitude of $\tilde \Lambda_0$, while those of $\tilde L$ are about one order of magnitude smaller.

\subsection{Detectability}
\label{sec:3C}
\noindent
Finally, let us comment on the prospects of detectability of subsolar PBH binaries when evolving in the vacuum or in an external environment. The faithfulness analysis reveals the degree to which a waveform model differs from another, such as comparing the presence or absence of the thin matter shell around each PBH. High faithfulness indicates that the models closely match each other across a broad frequency range, while low faithfulness may suggest discrepancies in the waveform templates, possibly leading to systematic and wrong interpretations of the considered GW event.

The result of the analysis is shown in Fig.~\ref{fig:Faith}.
The contour plots show the behaviour of the faithfulness $\mathcal{F}$ in terms of the PBH binary masses (left panel) or in terms of the thin matter shell parameters (right panel). In both plots one can appreciate that, for a large portion of the parameter space, the faithfulness takes values smaller than $\mathcal{F}_\text{\tiny th} \simeq 0.972$, allowing to potentially distinguish the two templates. The decreased values of ${\cal F}$ are due to the important dephasing induced by the environment for large $\epsilon$ and/or $\tilde L$. This mismatch would directly impact the efficacy of binary BH searches, based on template banks which neglect these effects. 

In the left panel, it is manifest that smaller values of $\mathcal{F}$ are reached for larger values of the binary total mass $M $. This happens since tidal effects start to be relevant at the 5PN order of the GW waveform, $\psi_\text{\tiny Tidal}  \propto (\pi M f)^{5/3} \tilde{\Lambda} (f)$, allowing for a smaller mismatch compared to naked PBH systems only when the binary is sufficiently light. This trend is expected to subsist until the ISCO frequency enters the observational band, which happens for binary total masses around $M_\odot$.

In the right panel, we show the behaviour of the faithfulness in terms of the matter shell parameters. As one expects, either small values of $\epsilon$ or $\tilde{L}$, both suppressing tidal effects in the waveform, induce no mismatch between the two templates. On the contrary, larger values of both parameters generate sizeable TLNs, giving rise to large dephasing, and therefore bigger differences between the models. Overall, a good portion of the parameter space experiences values of $\mathcal{F} \ll \mathcal{F}_\text{\tiny th}$, showing that neglecting environmental effects may lead to wrong interpretations of subsolar GW events.

\section{Conclusions}
\label{conclusions}
\noindent
We have explored the impact of environmental perturbations on the TLN of subsolar PBHs, a key observable in GW astronomy that could help to distinguish PBHs from other (non-)astrophysical compact objects. Our findings reveal that even small environmental disturbances, modeled similarly to the well-known ``flea on the elephant'' effect in the context of BH quasinormal modes, can induce a significant and detectable PBH TLN. This effect may complicate the use of tidal phenomena as a definitive tool for ruling out the PBH nature of a binary, as these disturbances can mimic the tidal deformability expected from other astrophysical objects.

However, we have shown that these perturbations induced by environmental effects, while initially complicating the TLN signal, are generally disrupted during the final stages of binary mergers due to tidal interactions. This disruption leaves a characteristic imprint on the gravitational waveform that current and future GW detectors could identify. Thus, despite the potential for environmental factors to generate large TLNs, the final signals from PBH mergers can still carry signatures indicative of their primordial origin.
As we have shown, the uncertainties on the cut-off frequency $f_\text{\tiny r}$ are much smaller than the ones on $\tilde \Lambda_0$.
Thus, if environmental effects become relevant, their time evolution would also be identified, and distinguished from intrinsic TLN of other compact objects. 

Our results emphasize the need for careful analysis of GW signals, considering possible environmental effects in the data analysis and  matched-filter searches, to accurately infer the nature of subsolar compact objects in binary mergers. This work advances our understanding of the complex interplay between environmental factors and GW observables and underscores the potential of current and future detectors to discover PBHs through subsolar searches.

We conclude with a few words of caution. 
First of all, we should stress that our modeling of the disruption of the environment is somehow simplified. As the matter shell is disturbed, energy is taken from the binary system. Therefore, one expects the binary evolution to be impacted through e.g. dynamical friction. Additionally, as the mass surrounding the object is dispersed, one may also expect signatures of the environmental effects to be captured by an effective time-dependent binary mass. While these phenomena are present in principle, we stress that the mass of the matter shell we assume here is suppressed by the combination $\epsilon \tilde L$ compared to the PBH one. We leave an estimation of these additional phenomena for future work. 
Eventually, the signal of a ``dirty'' PBH binary should be distinguished from the one of neutron stars, or other exotic compact objects, characterized by an intrinsic TLN. The different morphology characterizing the GWs from the two systems provides us with the opportunity to achieve this. In the first case, tidal effects are present in the early phase of inspiral, while leaving a ``naked'' PBH binary closer to the ISCO frequency. In the latter case, the tidal effects are always present, until disruption of the compact objects themselves~\cite{Crescimbeni:2024cwh}.


\section*{Acknowledgments}
\noindent
We thank Paolo Pani for the useful comments on the draft.
V.DL. thanks the University of Geneva and CERN for their warm hospitality during the realization of this project.
V.DL. is supported by funds provided by the Center for Particle Cosmology at the University of Pennsylvania. 
A.R. is supported by the Boninchi Foundation for the project “PBHs in the Era of gravitational wave Astronomy”.

\bibliography{draft}

\end{document}